# An sTGC Prototype Readout System for ATLAS New-Small-Wheel Upgrade

Peng Miao, Feng Li, ShengQuan Liu, ZhiLei Zhang, Tianru Geng, Xinxin Wang, Shuang Zhou, Ge Jin

*Abstract*–This paper presents a readout system designed for testing the prototype of Small-Strip Thin Gap Chamber (sTGC), which is one of the main detector technologies used for ATLAS New-Small-Wheel Upgrade. This readout system aims at testing one full-size sTGC quadruplet with cosmic muon triggers.

## I. INTRODUCTION

ATLAS[1] will replace the muon end-cap detectors, the so-called Small Wheel(SW), with the New Small Wheel (NSW) [2] in the Phase-I upgrade to enhance its high rate performance. Small-Strip Thin Gap Chambers (sTGCs), developing from the Thin Gap Chamber(TGC) technology but with much smaller strip pitch, have been selected as one of the main detector technologies to be used for the NSW. An sTGC quadruplet consists of four pad-wire-strip planes. To readout sTGC signals, two kinds of Front End Board (FEB)will be designed, pad Front End Board(pFEB) and strip Front End Board (sFEB). The pFEB with the maximum 192 channels is responsible for reading out pad and wire signals of each plane, while the sFEB with the maximum 512 channels is responsible for reading out strip signals of one gas-gap. This paper presents a readout system capable of testing one full-size sTGC quadruplet. It consists of 4 pFEBs and 4 sFEBs along with one specifically designed DAQ board. The FEBs use VMM3[3] ASIC for analog signal amplification and digitization. The DAQ board is able to configure and readout up to 8 FEBs through Gigabit Ethernet[4]. This readout system can be used to evaluate the functionality and performance of sTGC prototype, and help optimize the design of the final pFEB and sFEB.

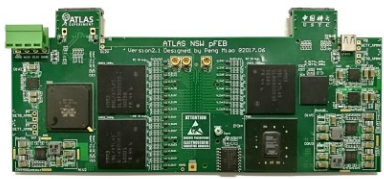

Manuscript received June 23, 2018. This work was supported by the National Natural Science Foundation of China under Grants 11461141010 and 11375179, and in part by "the Fundamental Research Funds for the Central Universities" under grant No. WK2360000005.

Peng Miao, Feng Li, ShengQuan Liu, ZhiLei Zhang, Tianru Geng, Xinxin Wang, Shuang Zhou, Ge Jin are with State Key Laboratory of Particle Detection and Electronics, University of Science and Technology of China, Hefei, Anhui 230026, P.R. of China (e-mail: mpmp@mail.ustc.edu.cn; peng.miao@cern.ch;phonelee@ustc.edu.cn;lsplsp@ustc.edu.cn;zzlei@mail.ustc.edu.cn;gudujian@mail.ustc.edu.cn;wxx10@mail.ustc.edu.cn;neo@mail.ustc.edu.cn; goldjin@ustc.edu.cn).

First author: Peng Miao, Corresponding author: Peng Miao.

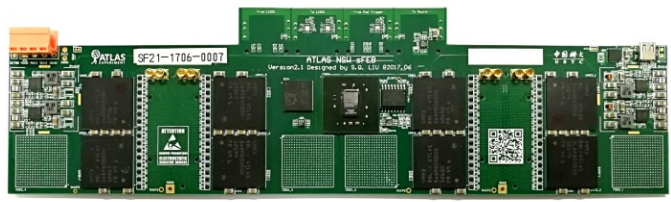

Fig. 1. The picture of the pFEB (version 2.1).

Fig. 2. The picture of the sFEB (version 2.1).

## II. READOUT SYSTEM SETUP

The simplified block diagram of the readout system for one sTGC quadruplet is illustrated in Fig.3.

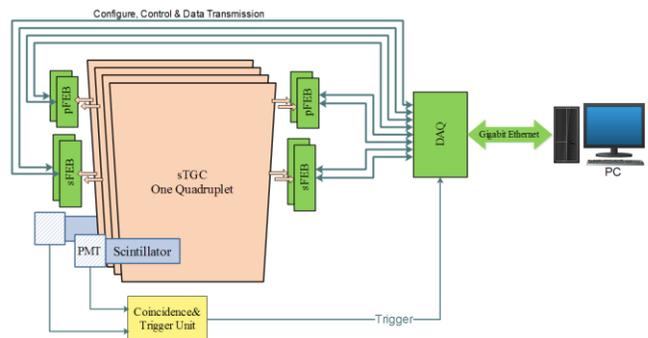

Fig. 3. Schematic of the readout system for sTGC prototype.

Each sTGC layer needs one set of pFEB and sFEB. The FEBs receive the charge signals from sTGC detectors through adapt-boards. The digitized amplitude value will be read out by a specifically designed DAQ board. This DAQ board is also responsible for configuring FEBs. The Graphical User Interface running on the host downloads command and configuration data to DAQ board and reads back packaged event data from DAQ board via Gigabit Ethernet.

The region of interest of sTGC detectors is sandwiched between two layers of scintillator arrays. And there is a PMT for capturing photons from each scintillator array. The Coincidence and Trigger Unit receiving the signals from two PMT generates trigger signals for the whole system when a muon goes through the sTGC quadruplet. Once triggered, the DAQ board will fan out the trigger signals to all FEBs through the control path.

## III. DESIGN OF DAQ BOARD

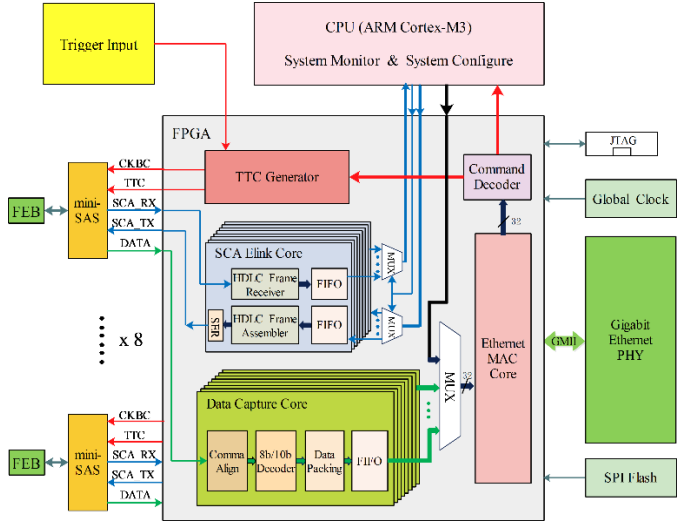
Fig. 4. Block diagram of the specifically designed DAQ board.

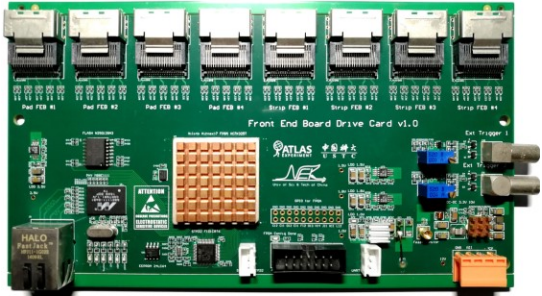
Fig. 5. Picture of the DAQ board.

Fig.4 shows the block diagram of the specifically designed DAQ board for this readout system. There are three relatively independent paths between FEBs and DAQ board, and they are control path, configuration path and data path.

The Timing-Trigger-Control (TTC) module is the key component of the control path. The TTC module has three main tasks: global clock fan-out, trigger signal fan-out and system synchronization signals fan-out.

The FEBs are all configured by the Slow Control Adapter ASIC (SCA), which is designed to provide several protocols for front-end configuration. The SCA chips on the front-end boards communicate with DAQ board through Elink, a full duplex serial interface. And the SCA channel command protocol is based on the HDLC standard. To deal with the complex HDLC protocol, a CPU/FPGA hybrid architecture is implemented, which significantly simplifies the FPGA firmware and makes the entire system much more reliable. The FPGA focuses on the high-speed HDLC frame reception and transmission, while the CPU concentrates on the complicated SCA commands parsing.

The triggered event data from FEBs is 8b/10b encoded. The data capture core in the data path performs 8b/10b decoding and data packaging. And then the Ethernet core uploads buffered data from eight separated FIFOs in a round-robin manner to PC for future analysis.

## IV. EXPERIMENTAL RESULTS

Fig.6 gives some typical analog signals from Front End Boards observed with oscilloscope. The trigger signal is from the Coincidence and Trigger Unit. And the right picture shows a four-pad-layer coincidence when a cosmic muon hits the sTGC quadruplet.

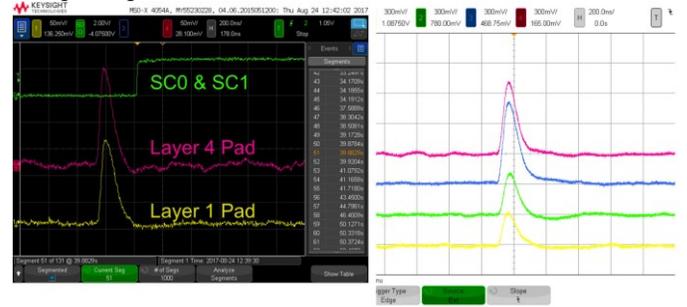
Fig. 6. Multi-layer coincidence with cosmic muon triggers. Two pad layers coincidence with trigger signal(left). And four pad layers coincidence(right).

The charge spectrum of sTGC pad and strip under the high voltage of 32kV is shown in Fig.7. The gain we use is 1mV/fC for sFEB and 3mV/fC for pFEB. The distribution is reasonable and the readout system can meet the design purpose for sTGC qualification.

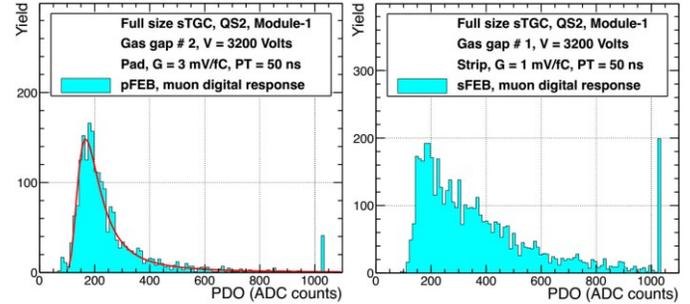
Fig. 7. Charge spectrum of Pad(left) and Strip(right), measured by this readout system.

## V. CONCLUSION

This readout system has been successfully used to test sTGC prototype at ShanDong University in China and Weizmann Institute of Science in Israel. And the experimental results show that this readout system satisfies the requirement of the sTGC prototype functionality and performance evaluation at each sTGC chamber production site.

ACKNOWLEDGMENT

We thank Ilia Ravinovich and Vladimir Smakhtin for performing the high science quality experiment, and the kindly help during my whole staying at WIS. We also thank Liang Guan and Siyuan Sun from University of Michigan for the help during the Test Beam experiment at CERN. And the authors would like to thank their colleagues from ATLAS Muon New Small Wheel Electronics collaboration for all the support and interesting discussions concerning this work.

REFERENCES

[1] ATLAS Collaboration, "The ATLAS experiment at the CERN large hadron collider," J. Instrum., vol. 3, p. S0800S, Aug. 2008.
[2] ATLAS Collaboration, "New small wheel technical design report," CERN, Geneva, Switzerland, Tech. Rep. CERN-LHCC-2013-006 or ATLAS-TDR-020, 1995.
[3] G. D. Geronimo et al., "VMM1—An ASIC for micropattern detectors," IEEE Trans. Nucl. Sci., vol. 60, no. 3, pp. 2314–2321, Jun. 2012.
[4] IEEE Standard for Ethernet, Section Four, IEEE Standard 802.3, 2012.